\documentclass[%
 reprint,
 amsmath,amssymb,
 aps,
]{revtex4-2}

\usepackage{graphicx}
\usepackage{dcolumn}
\usepackage{bm}
\usepackage{multirow}
\usepackage{xcolor}
\usepackage{hyperref}

\newcommand{\colored}[1]{\textcolor{black}{#1}}

\begin{document}

\preprint{APS/}

\title{Model-independent versus model-dependent interpretation of the SDSS-III BOSS power spectrum: Bridging the divide}

\author{ Samuel Brieden}
\email{sbrieden@icc.ub.edu}
\altaffiliation[Also at ]{Dept. de  F\'isica Qu\`antica i Astrof\'isica, Universitat de Barcelona, Mart\'i  i Franqu\`es 1, E-08028 Barcelona, Spain.}
\author{H\'ector Gil-Mar\'in}%
 \email{hectorgil@icc.ub.edu}
 \author{Licia Verde}
 \email{liciaverde@icc.ub.edu}
 \altaffiliation[Also at ] {ICREA, Pg. Llu\'is Companys 23, Barcelona, E-08010, Spain}
\affiliation{ICC, University of Barcelona, IEEC-UB, Mart\'i i Franqu\`es, 1, E-08028 Barcelona, Spain%
}
\date{\today}

\begin{abstract}
The traditional clustering analyses of galaxy redshift surveys compress the clustering data into a set of late-time physical variables 
in a model-independent way. This approach has recently been extended by an additional {\it shape variable}
encoding early-time physics information. We apply this new technique, {\it ShapeFit}, to  SDSS-III BOSS data and show
that it matches the constraining power of alternative, model-dependent approaches, which directly constrain the  model's parameters adopting a cosmological model {\it ab-initio}. {\it ShapeFit} is $\sim30$ times faster, model-independent, naturally splits early- and late-time variables, and enables a better control of observational systematics. 
\end{abstract}

\maketitle

\section{Large Scale Structure Clustering: Interpretation \label{sec:introduction}}

The traditional clustering analysis of large-scale structure (LSS) galaxy redshift surveys is done by compressing the power spectrum data products into physical variables in a largely model-independent way. These are the well known Alcock-Paczynski (AP) scaling factors  $\alpha_{\perp}$, $\alpha_{\parallel}$  \cite{1979Natur.281..358A} and the amplitude of velocity fluctuations, $f\sigma_8$ \cite{kaiser_clustering_1987,Percival:2008sh}. 
The AP scaling factors are obtained by observing the  standard ruler provided by the Baryon Acoustic Oscillation (BAO) feature.
The  amplitude of velocity fluctuations is obtained from the modulation of clustering amplitude in redshift space as function of the angle from the line-of-sight. This provides a powerful compression: from power spectrum multipoles as function of scale and redshift, to three quantities, the physical variables,  per redshift bin.   
\colored{These are the} physical variables that are then compared to theory predictions, within a given cosmological model, to constrain the numerical values of the model's parameters. The value of this classic approach lies in the fact that the model dependence is  introduced only at the very end of the process, leaving most of the analysis as model-independent as possible. In addition, this approach nicely disentangles information of  the late-time universe from that of the early-time universe, which is particularly valuable  for  going beyond simple parameter-fitting and pursuing ways to test the model and its underlying assumptions. 
It has a drawback, however: the compression is not lossless. Its target is robustness, but this comes at a cost.
\begin{figure*}
\centering
     \includegraphics[width=8cm]{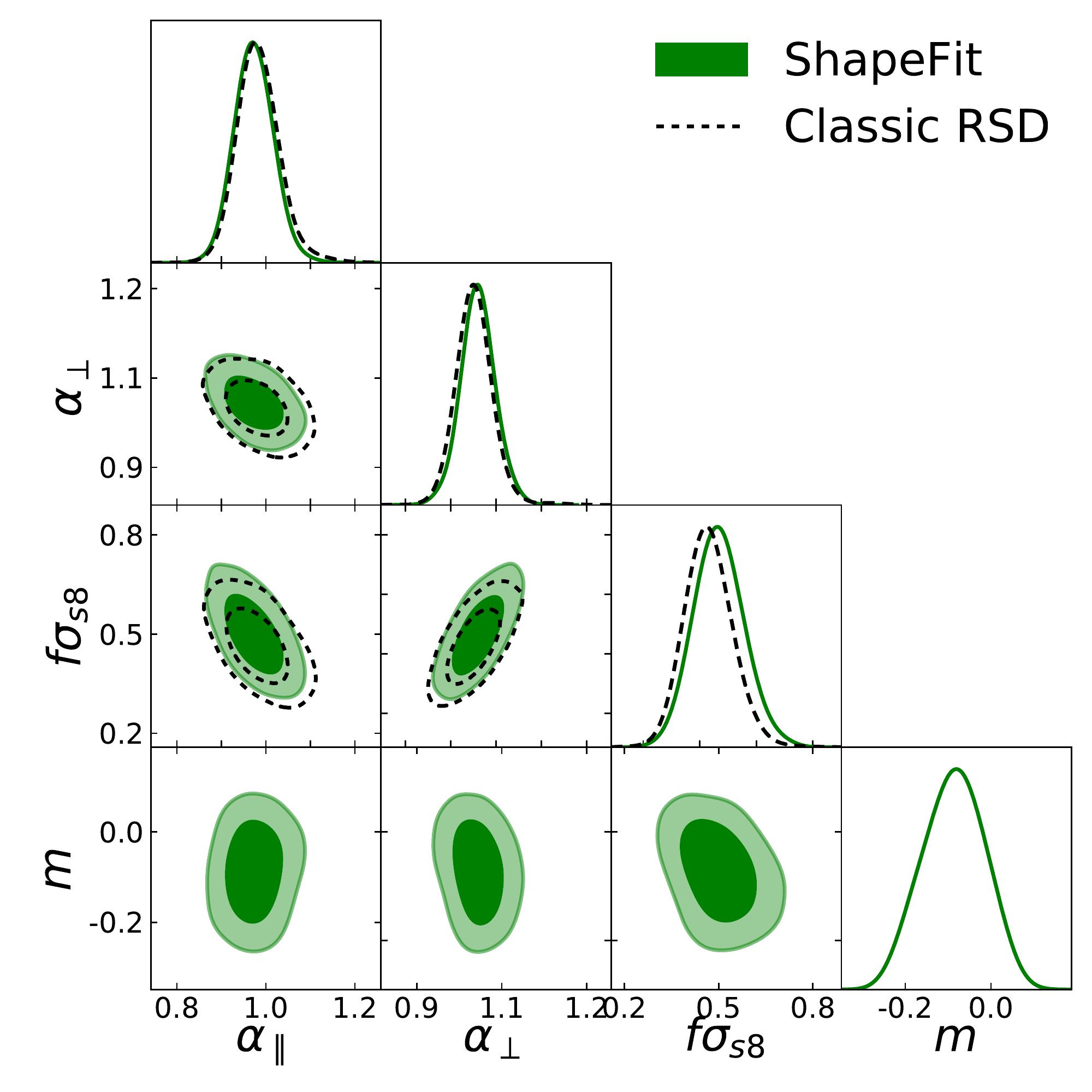}
    \includegraphics[width=8cm]{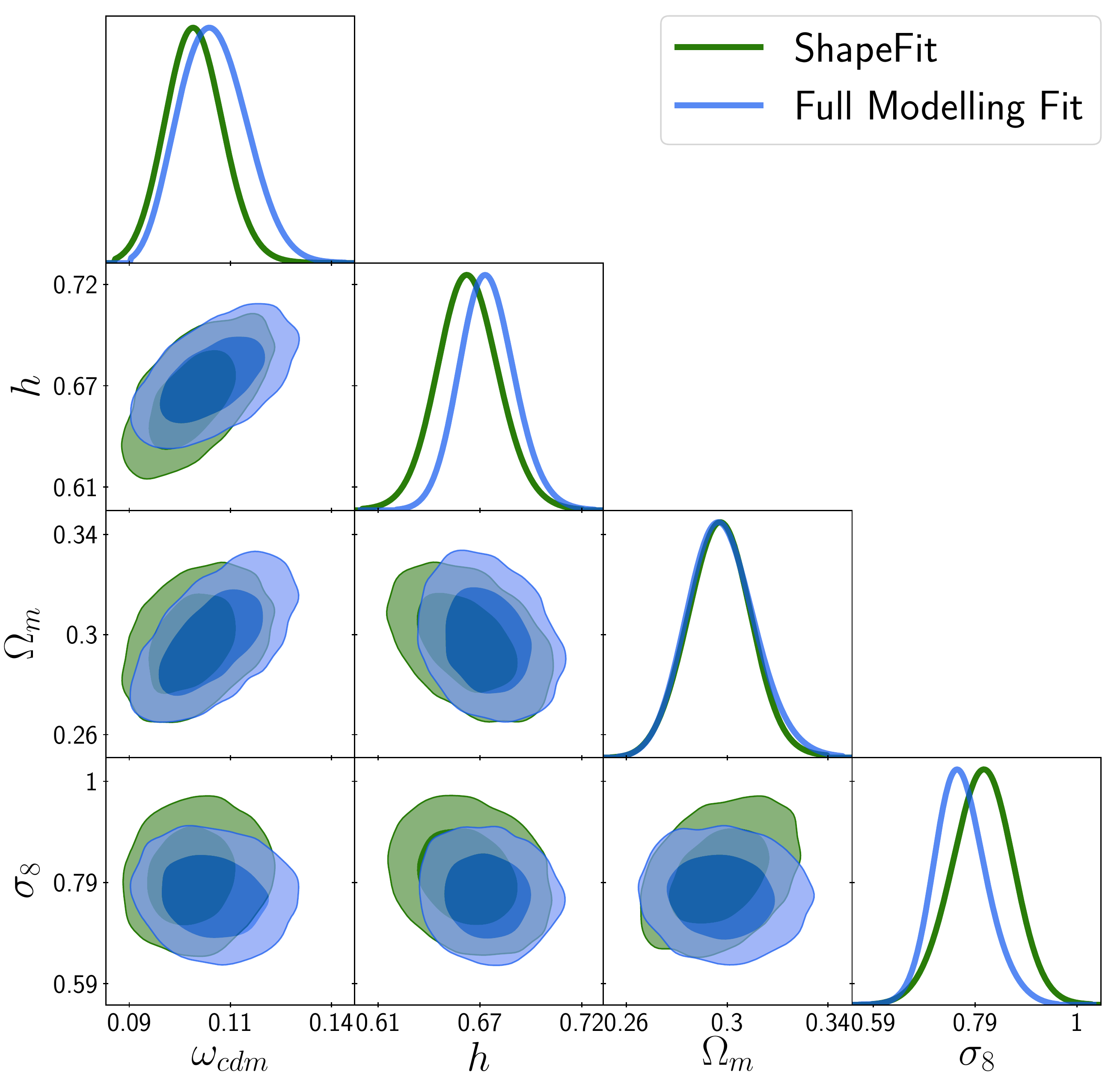}
\caption{{\it Left Panel}: compressed physical parameters posteriors derived from power spectra measurements of the BOSS high-redshift sample, $z_{\rm eff}=0.61$ (constraints from the low-redshift sample, show a  very similar behaviour). Black dashed contours display the classic RSD results, while novel {\it ShapeFit} results are shown in green. In both cases the 1-loop SPT theory has been used to model the monopole and quadrupole signals for $0.01\leq k\,[h{\rm Mpc}^{-1}]\leq 0.15$.  {\it Right Panel}: posteriors derived from low- and high-redshift samples of BOSS  using the same scale-cuts as in the left panel. The blue contours correspond to the FM approach when a flat-$\Lambda$CDM model (+BBN Gaussian prior on $\omega_b$) is directly fitted to the 224 power spectra multipoles bins, $P^{(\ell)}(k,z)$, using EFT to model the power spectrum. Conversely, green contours are drawn from the 8 compressed physical variables of {\it ShapeFit}, interpreted under the same cosmological model as for the blue contours.}
\label{fig:class-vs-shape_n_bossdr12}
\end{figure*}

This approach  is conceptually  different from the way in which, for example, Cosmic Microwave Background (CMB) data  are routinely analyzed. The CMB maps are compressed into  angular power spectra (as done for galaxy clustering), but then these are directly used to constrain the  values of the parameters of an adopted cosmological  model. The so called  ``physical parameters" for the CMB  were actually proposed in \cite{Kosowsky2002}. The original goal was to accelerate cosmological inference from CMB data, and some of these parameters are still employed to date for the computational speed-up they yield. But, in reality, the physical parameters capture phenomenological signatures of physical processes, and can then be interpreted {\it a posteriori} in terms of constraints on cosmological model parameters. 
The use of physical parameters in CMB analysis to produce model-independent constraints \cite{Wang:2006ts, 1912.04921} and further compress CMB observations is not mainstream, at least in part, for two reasons. The CMB gives us a snapshot of the photon-baryon plasma at recombination,  so  is located at a single cosmic epoch; moreover, CMB photons must cross the entire Universe from the last scattering surface to $z=0$, making it difficult to disentangle early-times physics signatures from late-times ones (but see \cite{audren12,verde17b}).  

The galaxy power spectrum can also be interpreted in a way completely analogous to the way the CMB is analyzed. The development of relatively fast (significantly faster than N-body simulations) modelling techniques for the non-linear galaxy power spectrum (e.g., Effective Field Theory, EFT) has made this ``full modeling" (FM) possible over the past couple of  years (\citep{Ivanov:2019pdj,damico} and references therein).
It  became quickly apparent that this newer approach produces much tighter constraints on cosmological parameters than the classic  (compressed-variables based) approach, if galaxy clustering is analyzed without external datasets, or strong external priors. On the other hand, in a joint CMB+LSS analysis (e.g., \cite{boss}) the two perform very similarly.

However, there is significant value in analyzing and interpreting galaxy clustering alone,  especially not in combination with early-time probes. 
Separate analyses of observations of disparate epochs of the Universe are key to shed light on recent cosmological tensions (e.g.,  \cite{hubbletension}), and propose explanations in terms of deviations from the standard cosmological model (e.g., \cite{DiValentino:2021izs}).

Until very recently, the extra signal responsible for  the spectacular improvement provided by the FM approach was not well understood. However \cite{ShapeFit} showed that a simple, one (phenomenological) parameter extension of the classic approach, {\it ShapeFit}, can capture most of this extra signal and provides the same statistical power within a flat-$\Lambda$CDM model.
The compression that {\it ShapeFit} provides is nearly lossless for models that are effectively described, or well approximated,  by $w$CDM-like models or simple variations of the CDM model at horizon scales at early times. 
While the classic approach (and {\it ShapeFit}) rely on a template for compression, it has been extensively demonstrated that the choice of the cosmological model necessary to create the template is unimportant, does not constitute a model prior and does not produce any significant systematic shifts under the correct interpretation of their physical variables \cite{eboss,Bernal_BAObias,ShapeFit}.

In the classic RSD fit, at a given redshift bin $z$, the full power spectrum multipoles, $P^{(\ell)}(k,z)$, are compressed in just three physical variables sensitive to late-time physics only. These are two background-level variables that describe the cosmic expansion in units of the standard ruler, $\alpha_\parallel(z)$ and $\alpha_\perp(z)$ (see section 2.4 of \citep{ShapeFit}); and a perturbation-level variable that describes structures growth, $f\sigma_8(z)$. The extra information that the classic RSD neglects (and that the FM captures) is related to the shape of the transfer function. In addition to a more appropriate definition of velocity fluctuations $f\sigma_{s8}$, {\it ShapeFit} introduces a new variable $m$ (see eqs. 3.5, 3.6 and 3.12 of \citep{ShapeFit} for definitions) which  captures very well the bulk of the  missing information. The physical interpretation of this $m$-variable is not any late-time physics phenomenon, but a series of early-time processes which modulate the broadband shape of the power spectrum (and the matter transfer function).
 
Hence, {\it ShapeFit} can be used to bridge the classic and FM approaches.  The connection lies on making explicit and enforcing (or removing) a key ``internal model prior” which ties together  early- and late-time compressed variables (see \cite{ShapeFit}).  While the compressed physical variables are model-independent, the internal model prior connects the signature of early-time physics on the clustering signal on large scales, to the standard ruler signature constraining the  late-time geometry and  the redshift space signature of  kinematics on the clustering.

\section{Application to SDSS-III BOSS data}
We employ the Luminous Red Galaxy (LRG) samples of the SDSS-III BOSS survey \cite{boss}, covering two non-overlapping redshift ranges: $0.2<z<0.5$ (effective redshift $0.38$),  containing  604,001 galaxies; and $0.5<z<0.75$ (effective redshift 0.61) containing  594,003 galaxies. As done in BOSS official papers, we treat these two redshift samples as uncorrelated. The effective volume traced by these two samples is $3.7\, {\rm Gpc}^3$ and $4.1\, \mathrm{Gpc}^3$, respectively, for  a total effective volume of $7.8\,{\rm Gpc}^3$.

This same data set yields very different cosmological constraints when it is analyzed using the classic approach or the FM fit (see  e.g., fig 2 of \citep{ShapeFit} grey  contours for classic RSD alone, orange when BAO post-reconstruction information is added,  blue for FM fit). 
Both approaches yield very similar constraints when combined with a CMB prior (e.g., Planck; see the right panel of fig. 2 in \citep{ShapeFit}), as this type of prior effectively fixes the early-time physics information enclosed in the  broadband shape. 

In what follows, parameter constraints are obtained with a standard Markov Chain Monte Carlo (MCMC) posterior sampling \cite{Brinckmann:2018cvx}. The modeling of the clustering signal follows \cite{ShapeFit,Ivanov:2019pdj} and employ the Boltzmann solver \cite{2011JCAP...07..034B} including the EFT extension from \cite{Chudaykin:2020aoj}.
The left panel of Fig.~\ref{fig:class-vs-shape_n_bossdr12} displays the constraints  on the late-time universe physical variables $\{\alpha_\parallel,\,\alpha_\perp,\,f\sigma_8\}$ obtained  by the classic RSD analysis (dashed black contours) and by {\it ShapeFit} analysis, with the extra early-time universe parameter $m$ (green contours), when both are applied to the high-redshift bin of BOSS. 

The constraints on the three late-time universe physical parameters are not significantly modified by the addition of $m$ as free extra variable, as $m$ is essentially uncorrelated with them. \colored{The small correlation between $m$ and, e.g., $f\sigma_8$ of $-0.3$ leads to only $5\%$ increase in errors.}

The posteriors of the left panel of Fig.~\ref{fig:class-vs-shape_n_bossdr12} have been obtained without any strong model assumption~\footnote{other than homogeneity, isotropy, and scale-independent growth. The reconstruction step assumes that gravity at mildly non-linear scales is well described by GR},   and hence  are easily interpretable within a wide set of cosmological models. This model-interpretation process essentially places `internal model priors' among the physical variables, connecting them with the internal parameters of the assumed model. This is shown by the green contours of the right panel of Fig.~\ref{fig:class-vs-shape_n_bossdr12}. The {\it ShapeFit} contours of the left panel (and additionally another set of four parameters at the low-redshift bin, $z_{\rm eff}=0.38$) are  interpreted within a flat-$\Lambda$CDM model with a  Gaussian big bang nucleosynthesis (BBN) prior  $\omega_b=0.02268\pm0.00038$ \cite{Adelberger:2010qa,Pisanti:2007hk,Ivanov:2019pdj,Schoneberg:2019wmt}; the resulting posteriors for $\{\omega_{cdm},\,\Omega_m,\,h,\,\sigma_8\}$ are drawn.  The constraints obtained  by directly fitting the $P^{(\ell)}(k,z)$ shape on the same range of scales under the FM approach using EFT theory to describe the $P(k)$ modelling are shown in blue. Note the spectacular agreement between both approaches, especially considering that  the green contours  are obtained from 
just 8 variables (the 4 physical variables, $\{\alpha_\parallel,\,\alpha_\perp,\,f\sigma_8,\,m\}$ at two redshift bins), while  blue contours are for 224 $P^{(\ell)}(k,z$) measurements (28 $k$-bins measurements for two multipoles, two redshift bins, and two galactic hemispheres). \colored{Another advantage of {\it ShapeFit} over the FM approach is computational time}. Once the compressed variables are extracted (since this step is model independent it only has to be done only once) the model-fitting is very fast: one model evaluation on a \colored{single-core is 8 times} faster than the \colored{FM run}. \colored{As} the cosmological interpretation of \textit{ShapeFit} parameters is done without any nuisance parameters and due to the much simpler likelihood surface, an MCMC  needs 5-10 times fewer sampled points than the FM method for the same level of convergence.
{\it ShapeFit} yields an overall speed-up factor  of 40-80.

 \section{the power of the shape variable}
      \begin{figure}
   \centering
    \includegraphics[width=\linewidth]{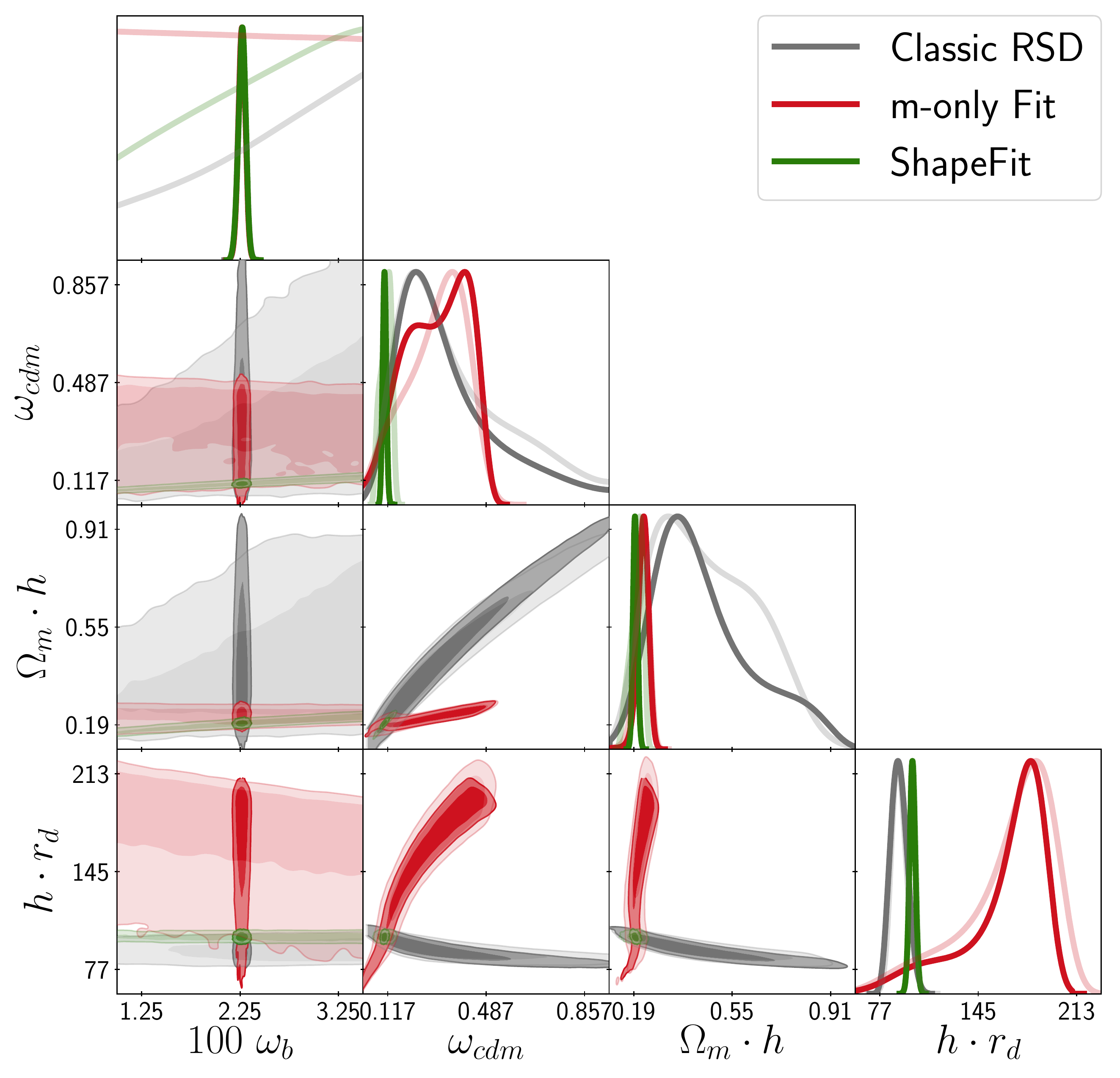}
 \caption{Interpretation within a flat-$\Lambda$CDM model with a Gaussian BBN prior on $\omega_\mathrm{b}$ (opaque contours) and without (transparent contours), of different physical variables constraints from the low- and high-redshift BOSS samples. Gray correspond to classic RSD analysis based on late-time  variables, $\{\alpha_\parallel(z),\,\alpha_\perp(z),\,f\sigma_8(z)\}$, red corresponds to the early-time shape variable $m(z)$ only, and their combination based on the $\Lambda$CDM internal model prior is shown in green.}
    \label{fig:mshape}
\end{figure}
 Fig.~\ref{fig:mshape} shows the cosmological constraints  for a standard flat-$\Lambda$CDM model,  obtained from the low- and high-redshift BOSS samples  using different sets of physical compressed variables. Gray contours arise from the classic RSD analysis using $\{\alpha_\parallel(z),\,\alpha_\perp(z),\,f\sigma_8(z)\}$, red contours from the {\it ShapeFit} analysis, but only using $m(z)$;  green contours represent the {\it ShapeFit} analysis using the full combination of 4 physical variables per redshift-bin (as for the right panel of Fig.~\ref{fig:class-vs-shape_n_bossdr12}).
 The transparent contours are for a broad uniform prior, $0.005<\omega_b<0.04$, the opaque contours for the Gaussian BBN prior.
 Note, that relaxing the prior does not significantly affect the 1D posteriors measured by the classic RSD and $m$-only fit, but broadens the {\it ShapeFit} result on $\Omega_\mathrm{m}h$ by a factor $\sim$ 2.5.

 The choice of parameters shown, $\{\Omega_m h,\,h r_s,\,\omega_{cdm},\omega_b\}$, highlights  the  complementary between the late- and the early-time physical variables.
The BAO signal naturally constrains $h r_s$\cite{bernal_trouble_2016}, while $m$ constrains $\Omega_m h$, as this variable is directly governing the shape of the matter transfer function via matter-radiation equality epoch. The relation between $m$ and $\Omega_m h$ is well approximated by the following fitting formula valid in the range $0.1<\Omega_m h<0.35$
\begin{align}
    \frac{\Omega_m h}{\Omega_m^\mathrm{ref} h^\mathrm{ref}} = 0.13 m^4 + 0.53 m^3 + 0.86 m^2 + m +1~.
\end{align}

Within a $\Lambda$CDM model, the purely late-time (uncalibrated) expansion history constrains the ratio $\alpha_\parallel/\alpha_\perp$ (also the relative isotropic signals among $z$-bins). This
can be used to measure $\Omega_m$, which is particularly well constrained  when low- and high-$z$ samples are combined (see fig. 5 of \citep{eboss}). 
In combination with the $\Omega_m h$ constraint provided by  $m$, it is thus possible to produce a measurement of $H_0$.
Note that, in spite of coming from galaxy clustering measurements, such measurement of $H_0$ {\it is not} arising only from late-time processes, but from a combination of early- and late-time universe physics. Following this procedure we use  the $\Omega_m h$ measurement from the $m$-only analysis of BOSS LRGs data for $0.2\leq z \leq 0.75$ (red contours of Fig.~\ref{fig:mshape}, $\Omega_mh=0.220^{+0.029}_{-0.019}$, without  the BBN prior on $\omega_b$),  with the $\Omega_m$ constraint from the uncalibrated BAO of the full BOSS+eBOSS sample: $\Omega_m=0.299\pm 0.016$, see table 4 of \citep{eboss}, which includes clustering measurements of low-redshift galaxies, LRGs, Emission Line Galaxies, quasars \colored{and Lyman-$\alpha$ emission lines} \colored{(or $\Omega_m=0.330\pm 0.037$ without Lyman-$\alpha$)}.
The $\Omega_m h$  and $\Omega_m$ measurements are considered uncorrelated as they come from different physical effects and different scales ($m$ is \colored{almost} uncorrelated with standard BAO variables, left panel in Fig.~\ref{fig:class-vs-shape_n_bossdr12}).
We find $H_0=73.6^{+10.5}_{-7.5}$, \colored{(or $H_0=66.7^{+12.1}_{-10.1}$ without Lyman-$\alpha$, where the change is solely driven by the determination of $\Omega_\mathrm{m}$)},
independent of any prior on $\omega_b$, or the absolute length of the BAO standard ruler. 
 \begin{figure}
   \centering
     \includegraphics[scale=0.23]{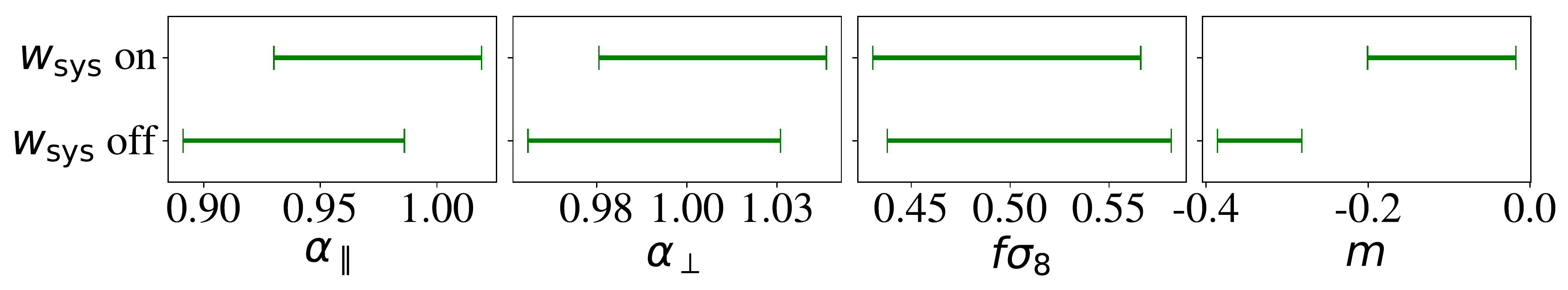}
    \includegraphics[scale=0.45]{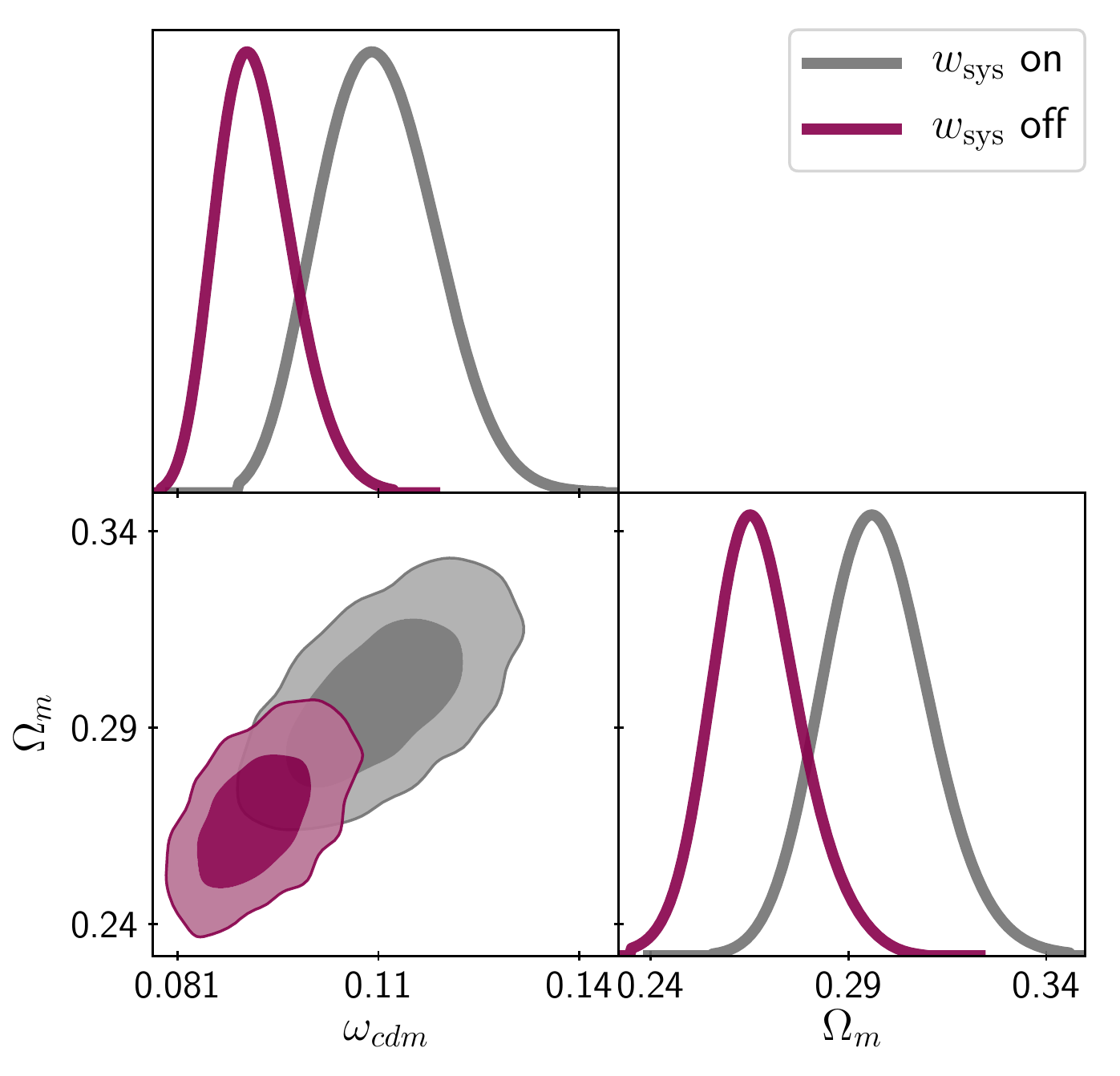}

 \caption{Effect of turning on and off the imaging systematic weights of BOSS data: for {\it ShapeFit} in its compressed set of physical variables (upper panels); and for the FM fit in the $\Omega_m-\omega_{cdm}$ plane (lower panel). For {\it ShapeFit} $f\sigma_8$ and  $\alpha_{\parallel,\,\perp}$ are barely affected by this correction, whereas $m$ absorbs most of the effect;  for FM fit, $\omega_{cdm}$ and $\Omega_m$ are significantly biased.}
    \label{fig:sys}
\end{figure}
We also report the value of $H_0$  obtained from applying {\it ShapeFit} to the LRG sample in combination of a BBN prior on $\omega_b$ (this is what is shown in the right panel of Fig.~\ref{fig:class-vs-shape_n_bossdr12}): $H_0=66.0^{+2.0}_{-1.7}$.

To quantify the impact of the known imaging systematics 
on cosmological constraints we repeat the above analysis by setting the systematic weights to unity in the BOSS catalogues (i.e., no correction for imaging systematic effects).
As shown in Fig.~\ref{fig:sys}
the scaling parameters and  $f\sigma_8$ are left largely unchanged while $m$ is affected by a shift of about 2.4 $\sigma$. Not unsurprisingly, $m$ ``absorbs"  systematic effects such as seeing, completeness or extinction angular dependencies: 
late-time physics constraints from clustering measurements are  significantly more robust than early-time physics constraints.   
      
 \begin{figure*}
   \centering
    \includegraphics[scale=0.35]{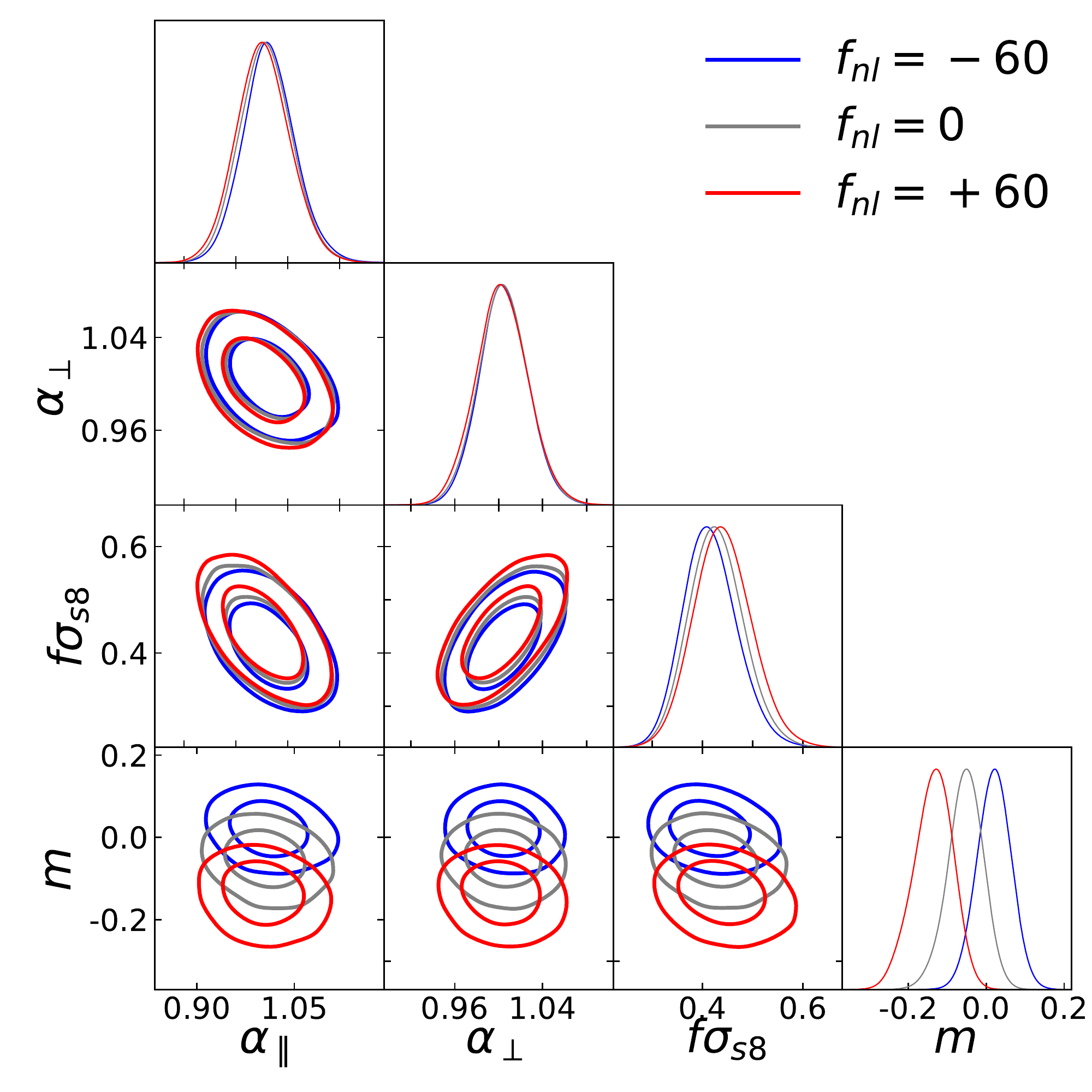}
    \includegraphics[scale=0.55]{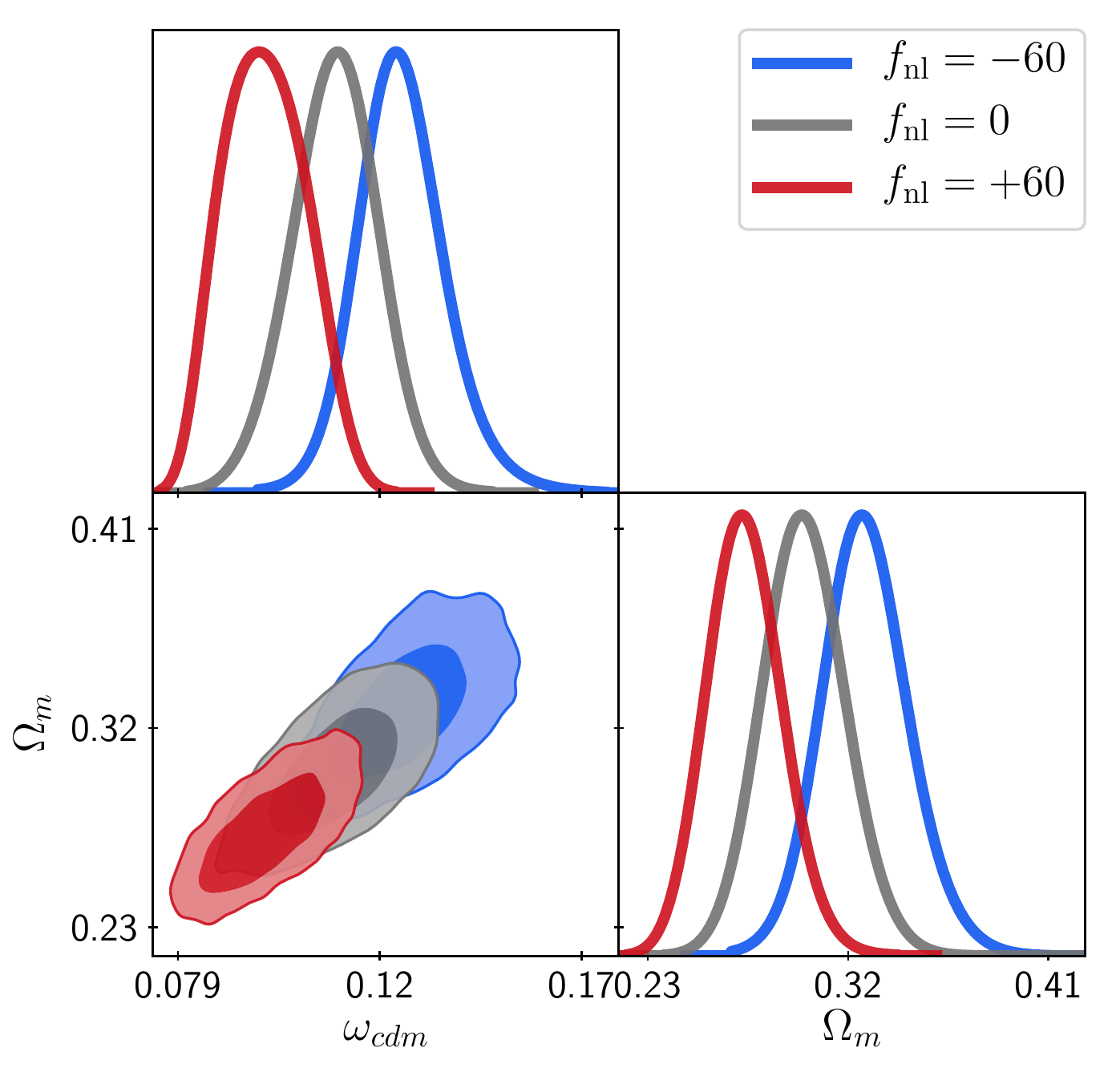}
 \caption{Systematic bias caused by ignoring in the modelling a $f_{\rm NL}$-signal which is present in the data-vector. In this case we have imprinted a mock $f_{\rm NL}=\pm 60$ signal, which is represented by red and blue contours. For {\it ShapeFit} (left panel) this systematic effect only impacts the shape parameter $m$ leaving $f\sigma_8$ and the scaling parameters unaffected. For the the FM fit (right panel) it biases both $\Omega_m$ and $\omega_{cdm}$.}
    \label{fig:fnl}
\end{figure*}
      
Finally, the advantage offered by a model-independent approach like {\it ShapeFit} can be appreciated  by devising a situation where the internal consistency check fails.

It is well known that a primordial non-Gaussianity of the local type induces a scale-dependent bias in the clustering of biased tracers, which is important at very large scales \cite{Dalal:2007cu,Matarrese:2008nc}. This scale-dependent bias correction is proportional to the linear bias, the non-Gaussianity parameter $f_{\rm NL}$ and has a scale dependence $\sim 1/k^2$, hence a leakage of this signal into $m$ can be expected.
We forecast the performance of {\it ShapeFit} and FM by generating mock power spectrum monopole and quadrupole signals according to 2-loop resummed perturbation theory, and analyzing it as done for the BOSS NGC $0.5\leq z\leq0.75$ data with the same covariance matrix. For choices of bias parameters consistent with the bias of BOSS galaxies ($b\sim2.2$), the effective redshift of BOSS and including only $k>0.01 h$ Mpc$^{-1}$, we find that a $f_{\rm NL}=\pm 60$ induces a change in $m$ of $\Delta m= \mp 0.08$ or, in general \colored{(linear response validated also for intermediate values)}, $\Delta m=  -0.0013 f_{\rm NL}$, leaving all other physical parameters unaffected. This is shown in the left panel of Fig.~\ref{fig:fnl}: the presence of non-zero $f_{\rm NL}$ does not bias the recovery and cosmological interpretation of \colored{$\alpha_\parallel$, $\alpha_\perp$} and $f\sigma_8$.  

The right panel of Fig.~\ref{fig:fnl} shows the effect on $\omega_\mathrm{cdm}$ and $\Omega_\mathrm{m}$ (other cosmological parameters are unaffected) of applying the FM pipeline to the same datasets containing a primordial non-Gaussian signal. Since the FM analysis avoids the compression step, the bias induced by $f_{\rm NL}$ directly propagates into model parameters, without the possibility to diagnose where the signal actually comes from, as it is the case in the {\it ShapeFit} approach.
This indicates that in the presence of non-zero $f_{\rm NL}$, a FM analysis assuming Gaussian initial conditions would recover biased results  for $\Omega_m$ and $\omega_{cdm}$. 
The difference in \colored{$\chi^2$-estimation} between the fit for $f_{\rm NL}=0$ and that for $f_{\rm NL}=60$ is $\colored{\Delta} \chi^2=5$ for FM (54 data points, 10 parameters), indicating that a "goodness-of-fit" test relying on $\chi^2$ values would not be enough to signal any issue.

It is important to note that the scale-dependent bias effect of $f_{\rm NL}$ is usually considered negligible at scales $k>0.03 h$  Mpc$^{-1}$, hence the leakage of $f_{\rm NL}$ on $m$ for {\it ShapeFit} and the shift in $\omega_{cdm}$ and $\Omega_m$ for FM,  is expected to become significantly more important for surveys volumes that probe scales $k<0.01 h$ Mpc$^{-1}$ not included here.    

\section{conclusions}
For the BOSS dataset the  shape-parameter efficiently captures the extra information that FM approaches deliver. {\it ShapeFit}, by  working in terms the compressed variables,  has essentially three main advantages over FM.

 {\bf Model-independence and computing time}. Once  constraints on the physical variables are obtained  they can be interpreted within multiple cosmological models at minimum computational cost. On the other hand, the full modelling approach requires to re-run the full analysis for each new choice of cosmology.
 
 {\bf Physical Insight}. The physical variables are naturally  directly related to specific physical processes that happen in the Universe at different epochs. The scaling factors and the growth of  perturbations are sensitive only to the late-time physics of the Universe. The shape parameter captures the shape of the power spectrum on large-scales ($\sim$ to the horizon size at $z \gtrsim 1000$) which contains signatures of early-time physics.
For a given cosmological model the early- and late-time effects are intrinsically related, which {\it i)} sets an internal model-prior implicit in the full model approach but made explicit in the {\it ShapeFit}; {\it ii)} the early- and late-time physical variables can be used to perform a powerful consistency test of the cosmological model. 
      
 {\bf Systematics control.} The {\it ShapeFit}  analysis (as well as classic) naturally separates the cosmological information into variables which have very different systematic budgets. The BAO-inferred signal has been shown to be extremely robust to theoretical and observing systematics, with a conservative error budget for state-of-the art measurements of $\lesssim 1\%$  \citep{baosys}. 
 The amplitude of velocity fluctuation can suffer from imaging and spectroscopic systematics if these are not exquisitely taken into account. The current estimate for this systematic budget is $\simeq2\%$ \cite{fs8sys}.
 The shape parameter can severely suffer from observational large-scale systematics (e.g.,  extinction, seeing, completeness). For BOSS data we quantify that the known imaging systematic produces a $\sim 2.4 \sigma$ shift in $m$ if not corrected.
 On the other hand, it absorbs non-standard early-universe physics signals and prevents them to leak into and bias the determination of late-time parameters shaping the expansion/growth history. 

We envision that the connection between the physical variables proposed by {\it ShapeFit} and the full modeling approach will provide a transparent bridge between model-independent and model-dependent interpretation of forthcoming galaxy redshift surveys and a direct physical understanding of their clustering results.  

\paragraph*{Acknowledgements.}
\footnotesize{
H.G-M. and S.B. acknowledge the support from ‘la Caixa’ Foundation (ID100010434) with code LCF/BQ/PI18/11630024.
L.V. acknowledges support of European Unions Horizon 2020 research and innovation programme ERC (BePreSySe, grant agreement 725327). Funding for this work was partially provided by the Spanish MINECO under projects PGC2018-098866-B-I00 FEDER-EU. 
Funding for SDSS-III has been provided by the Alfred P. Sloan Foundation, the Participating Institutions, the National Science Foundation, and the U.S. Department of Energy Office of Science. The SDSS-III web site is http://www.sdss3.org/.}

%
\end{document}